\documentclass[journal,onecolumn]{IEEEtran}
\usepackage{amsmath,amsfonts}
\usepackage{algorithmic}
\usepackage{algorithm}
\usepackage{array}
\usepackage[caption=false,font=normalsize,labelfont=sf,textfont=sf]{subfig}
\usepackage{textcomp}
\usepackage{stfloats}
\usepackage{url}
\usepackage{verbatim}
\usepackage{graphicx}
\usepackage{cite}
\usepackage{siunitx}
\usepackage{xcolor}
\usepackage{cancel}
\usepackage{makecell}
\usepackage{multirow}
\usepackage{mathrsfs}
\usepackage{booktabs} 
\usepackage{xspace}
\usepackage{hyperref}
\usepackage{pdfpages}

\newcommand{\V}[1]{\ensuremath{\boldsymbol{#1}}}

\newcommand{\D}[1]{\ensuremath{#1}}

\newcommand{\modulus}{ \ensuremath{\V a}}

\newcommand{\phase}{ \ensuremath{\V\phi}}
\newcommand{\transmittance}{\ensuremath{ t}}
\newcommand{\hologram}{\ensuremath{{\V{d}}}}
\newcommand{\model}{\ensuremath{\V{m}}}

\newcommand{\noise}{\ensuremath{\D{\eta}}}

\newcommand{\M}[1]{\ensuremath{\boldsymbol{\mathrm{#1}}}} 
\newcommand{\zprop}{\ensuremath{z}} 
 
\newcommand{\Hz}{\M{H}_{\zprop}\xspace} 
\newcommand{\Hbz}{\M{H}_{-\zprop}\xspace} 
\newcommand{\Hbdz}{\ensuremath{\M{\underline{H}}_{-2\zprop}}\xspace} 
\newcommand{\deltaobjtransmcplx}{\ensuremath{\V{o}}\xspace}

\begin{document}

\title{Physics-based self-supervised learning of a deep network for single-shot in-line hologram reconstruction}

\author{Dylan Brault, Félix Riedel, Corinne Fournier, Thomas Olivier, Loïc Denis 
\thanks{D. Brault is with PSI Center for Life Sciences, 5232 Villigen PSI, Switzerland (e-mail: \href{mailto:dylan.brault@telecom-st-etienne.fr}{dylan.brault@telecom-st-etienne.fr}).
F. Riedel, C. Fournier, T. Olivier, and L. Denis are with Université Jean Monnet Saint-Etienne, CNRS, Institut d Optique Graduate School, Laboratoire Hubert Curien UMR 5516, F-42023, SAINT-ETIENNE, France.\\
This work was supported by Agence Nationale de la Recherche (ANR), project ATICS (ANR-23-CE51-0023), and by the NIH BRAIN CONNECTS program of the National Institutes of Health under award number 5U01NS132317 and by the SERI-funded ERC Starting Grant \#MB22.00042.\\ 
(Dylan Brault and Félix Riedel are co-first authors.)
(Corresponding author: Dylan Brault.)
 $\copyright$ 2026 IEEE.  Personal use of this material is permitted.  Permission from IEEE must be obtained for all other uses, in any current or future media, including reprinting/republishing this material for advertising or promotional purposes, creating new collective works, for resale or redistribution to servers or lists, or reuse of any copyrighted component of this work in other works.}}

\markboth{IEEE TRANSACTIONS ON COMPUTATIONAL IMAGING, March~2026}%
{Shell \MakeLowercase{\textit{et al.}}: A Sample Article Using IEEEtran.cls for IEEE Journals}

\maketitle

\begin{abstract}
Digital in-line holographic microscopy is a computational imaging method useful for characterizing the refractive properties of a sample, \textit{i.e.} the phase shift and absorption.
This indirect measurement technique captures a diffraction pattern and uses reconstruction algorithms to retrieve the optical properties of the sample.
Since only the intensity of the diffracted wave is recorded on the sensor, this inversion is not trivial, and simple backward propagation leads to artifacts known in optics as the ``twin-image''.
With advances in deep learning, various algorithms have been developed for the reconstruction of in-line holograms, providing computationally efficient alternatives to iterative algorithms.
These algorithms rely either on supervised learning, which requires ground truth knowledge, or physics-based self-supervised algorithms that require additional information, like phase diversity, but require multiple holograms for inference.
This paper introduces a new self-supervised physics-based deep learning strategy that leverages phase diversity during training and then reconstructs sample's transmission function from a single in-line hologram during inference.
We introduce five datasets of simulated and experimental in-line holograms of beads and bacteria. The proposed method produces accurate quantitative reconstructions similar or even more accurate than those obtained by regularized inversion while reducing the computational time by a factor of 1000.
\end{abstract}

\begin{IEEEkeywords}
Phase retrieval, self-supervised deep learning, physics-based deep learning, digital in line holography.
\end{IEEEkeywords}

\section{Introduction}

Characterizing the refractive properties of a sample is a widely studied topic in the computational imaging field.
Phase imaging techniques such as holography, ptychography and diffractive tomography have been used to image transparent samples with electromagnetic waves (optical~\cite{huang1971digital, yeom2006real,kamilov2016optical, ren2019fringe,priscoli2021neuroblastoma}, X-ray~\cite{miao2011coherent} and terahertz~\cite{choporova2015classical}) and electron microscopy~\cite{gabor1948new,ophus2023quantitative,sanchez2025phase}.
In these configurations, the phase of the wave in the sensor plane is not directly measured as the sensors only record intensity.
Retrieving the phase-shift introduced by the sample, \textit{i.e.} its refractive properties, from these intensity measurements involves solving an ill-posed inverse problem.

Digital in-line holographic microscopy (DIHM) is a computational imaging technique that relies on two steps, the acquisition of a hologram and its numerical reconstruction~\cite{huang1971digital,goodman2005introduction}.
The digital hologram is obtained by illuminating a sample with a (partially) coherent incident wave and recording the intensity of the diffraction pattern produced by the sample on a camera.
The hologram can also be seen as an out-of-focus measurement of the sample.
Reconstruction algorithms aim at retrieving the complex-valued transmission function of the sample, which describes how the sample absorbs and shifts the phase of the incident wave.
Because only the intensity of the wave is recorded on the sensor (and not the phase), simple back-propagation reconstructions are hampered by the intrinsic ``twin-image'' issue, wherein a superimposed, out-of-focus virtual image of the object corrupts the reconstructed image of the sample.

Many advanced algorithms have been suggested to retrieve the sample's phase-shift by suppressing the twin-image artifacts.
These approaches, grounded in physical modeling of diffraction, treat the reconstruction task as an optimization problem.
Iterative phase retrieval (PR) algorithms are among the most popular.
However, they are computationally expensive as they rely on multiple propagations of the wavefront.

The Gerchberg-Saxton (GS) algorithm~\cite{gerchberg1972practical} is a widely used PR algorithm based on phase diversity of two holograms of the same specimen, taken at two propagation distances ($z$ and $z+\delta z$).
The GS algorithm iteratively propagates the wavefront between the two planes, imposing the measured intensity as a constraint, at each plane, while letting the phase evolve through the iterations.
After convergence, the reconstructed wavefront is back-propagated to the object plane, yielding the estimation of the in-focus transmission function.

In many experimental contexts, like live imaging, it is difficult to acquire two holograms at two different propagation distances.
Iterative methods using a single hologram, like error-reduction algorithms, have been proposed for these cases.
These methods, initially introduced by Fienup in~\cite{fienup1978reconstruction,fienup1982phase,guizar2012understanding} and subsequently applied to holography by Latychevskaia~\cite{latychevskaia2007solution}, leverage prior knowledge of the sample as optimization constraints.
As an example, the sample's non-emissivity prior can be imposed using a positivity constraint on the object transmission function.

These iterative reconstruction techniques are generalized by the regularized inverse problems approaches (IPA)~\cite{fienup1982phase, momey2019fienup}.
Within this framework, the reconstruction task implies minimizing a loss function combining two terms: a term defining how closely the data matches the theoretical image formation model, and a term capturing the prior knowledge of the sample.
This minimization problem is equivalent to a classical maximum \textit{a posteriori} (MAP) estimation where the data fidelity term corresponds to the negative log-likelihood, and the regularization term is the negative log prior.
These more comprehensive approaches allow for greater flexibility in the selection of object priors and improve the accuracy of DIHM reconstruction~\cite{sotthivirat2004penalized, denis2009inline, Lee_07, soulez2007inverse, brault_multispectral_2023}.
Nevertheless, they are computationally expensive and depend either on closed-form regularization priors, or on learned regularization priors~\cite{venkatakrishnan2013plug,chan2016plug,romano2017little} which require ground truth data for training.

In recent years, the rise of deep learning approaches has also impacted the field of inverse problems in imaging~\cite{ongie2020deep-3ad}.
After a neural network has been trained to solve an image reconstruction task, inference on new images can be very fast.
They have been applied successfully to phase reconstruction problems, including digital in-line holography applications~\cite{situ2022deep, wang2024use}.
Among these approaches, supervised training of a deep neural network is the easiest way to train a network to reconstruct an image when pairs of inputs (data and ground truth image) are available.
The network is trained by minimizing the discrepancy between the reconstructed transmission function and the ground truth image.
First attempts to reconstruct holograms with deep-learning considered a simple approach in which holograms were fed directly to the deep neural network~\cite{sinha2017lensless-701,wang2018eholonet-119}.

Closer to classical iterative reconstruction approaches, unrolled methods incorporate the physics of the problem by mimicking the iterations of a reconstruction algorithm~\cite{gregor2010learning}.
In this framework, each block of the network corresponds to a single iteration of the unrolled algorithm~\cite{metzler2018prdeep,zhang2021physicsbased-aba,baoshun2022dualprnet,yang2023hionet}.
For example, to reproduce a proximal gradient descent, each layer of the unrolled network can be broken down into a gradient descent step based on the image formation model, followed by a small regularizer network that acts as a proximal operator.
As the regularizer network has many degrees of freedom, it can provide more realistic priors than a handcrafted one.

Exploiting the high number of degrees of freedom of neural networks, generative adversarial neural networks have also been exploited to solve inverse problems in DIHM~\cite{yin2019digital-6ea,zhang2021phasegan-e55}.
In this framework, a generator is first trained to reproduce images that follow the distribution of the objects (and hence fool the discriminator).
The network is then used as a strong prior when reconstructing the transmission function images.
The reconstruction task is formulated as the retrieval of the latent vector that, once fed to the network, generates an image that is close to the data according to the image formation model.

These approaches have been applied successfully to digital in-line hologram reconstruction, but they rely heavily on ground truth datasets.
Such paired datasets of diffracting objects and holograms were obtained by using an SLM to display phase patterns coming from publicly available image dataset, such as Faces-LFW, ImageNet, and MNIST in the work of ~\cite{wang2018eholonet-119}.
However, these images hardly represent the samples encountered in real-world applications of holographic imaging.
Alternatively, but at a high computational cost, experimental holograms can be reconstructed using other existing phase reconstruction techniques to provide surrogate ground truth data~\cite{rivenson2018phase, wang2019net,moon2020noise,chen2022fourier-740}.

To avoid the acquisition/reconstruction of ground truth datasets, self-supervised approaches are being increasingly considered.

As the architecture of convolutional neural networks acts as an implicit prior that favors natural images outputs, it was proposed to use a neural network as a regularization prior.
This principle relates to ``Deep Image Prior''-based methods where a network fed by a random image is trained to reconstruct an image~\cite{li2020deep-38b,niknam2021holographic-d7e,galande2023untrained-b57}.
Similarly to inverse problems approaches, physical modeling can also be included in the loss function of these networks, as proposed by Wang et \textit{al.} \cite{wang2020phase-5b6}.
In these approaches, the reconstruction aims at estimating the parameters of a neural network that, once fed with a given random vector, provides the best reconstruction according to the physics-based loss function.
Neural networks have a large number of degrees of freedom, allowing more realistic reconstructions than handcrafted prior, such as sparsity and total variation (TV). However, interpreting the learned prior is more challenging due to the high number of degrees of freedom.
Moreover, these approaches often rely on early stopping of the training because there is a risk of overfitting the data if the training goes on for too many iterations.
A major drawback is that a different network needs to be trained for each input image, leading to a high computational-time per reconstruction.
There is no need for a training dataset, but the downside is that the network does not leverage common features that could be extracted from multiple data.

To exploit both the capabilities of deep neural networks to produce high-quality reconstructions while leveraging multiple data, physics can be incorporated directly in the training loss function.
Such a self-supervised approach is close to the inverse problem one.
Yet, instead of estimating a single reconstruction, the network learns a reconstruction operator that can be applied to many images.

To the best of our knowledge, these physics-based approaches have only been applied to reconstruct the transmission function from multiple out-of-focus measurements in DIHM~\cite{li2022physicsenhanced-6e5,huang2023self}. Such a setting can be experimentally prohibitive.
Training a neural network that can reconstruct the transmission function from a single measurement has a much stronger practical interest.
The contributions of this paper are the following :
\begin{itemize}
    \item introducing a \emph{self-supervised training} of the network, relying on phase diversity (pairs of out-of-focus measurements) and physics (hologram formation model);
    \item providing a \emph{fast} reconstruction method from a single-shot hologram, using a deep neural network;
 \item obtaining \emph{quantitative} phase reconstructions that match the reconstructions achieved by state-of-the-art methods based on multiple holograms (pairs of holograms recorded at two depths);
 \item releasing five \emph{open datasets}of simulated and experimental holograms of spherical objects and bacteria to promote further research on the topic and allow for reproducible research.
\end{itemize}

The paper is organized as follows.
Section \ref{sec:proposedApproach} details the principle of our approach.
Section \ref{sec:Recons_holo_simu} and \ref{sec:Recons_holo_expe} validate this approach on simulated and experimental DIHM holograms by providing comparison with state-of-the-art reconstruction algorithms.

\section{Proposed self-supervised method for single-shot hologram reconstruction}\label{sec:proposedApproach}

\begin{figure*}[t]
    \centering
    \includegraphics[width=0.9\linewidth]{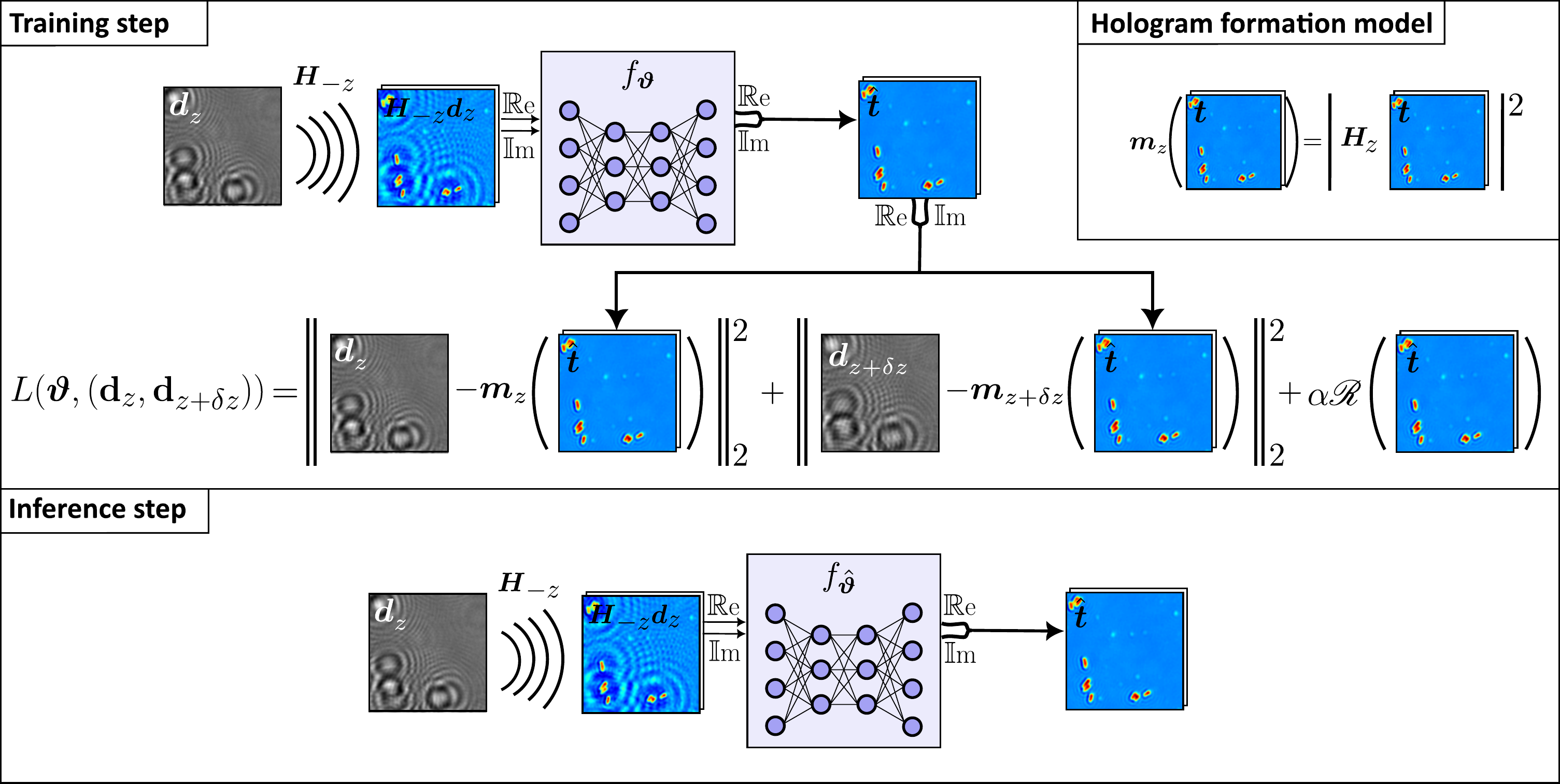}
    \caption{Framework of the proposed self-supervised approach. For illustration purpose, the vectors $\V \transmittance$, $\hologram_z$, $\Hbz \V{\hologram_z}$ are represented as 2D images.}
    \label{fig:principle}
\end{figure*}

This section introduces the image formation model of DIHM and discusses the ``twin-image'' problem to lay the groundwork for our approach.
Then, we present the principle of our self-supervised phase diversity and physics-based deep learning approach.
The training step uses holograms defocused at various distances as well as image formation models.
The inference step, uses a single hologram for the reconstruction.
To clarify the terminology, we describe our technique as self-supervised as it does not require any ground truth for training; however, additional measurement is necessary for the training step.
The principle of our approach is outlined in Fig.\ref{fig:principle}.

\subsection{Image formation model}\label{sec:ImageFormationModel}
The optical properties of a biological sample on a glass slide can be described by the complex-valued transmittance function in the object plane.
In discrete form, this two-dimensional (2D) transmission function is represented by the vector ~$\V \transmittance\in \mathbb{C}^N$, where $N$ is the number of pixels.
Its modulus $\V \lvert \V{\transmittance} \rvert = \modulus\in \mathbb{R}_+^N$ characterizes the absorption of the sample, while its phase shift $\V \phase\in [-\pi,\pi[^N$ accounts for the delay induced on the incident wavefront by the sample:
\begin{equation}\label{eq:transmissionFunction}
    \V{\transmittance}=\V{\modulus} \exp \left[i \V{\phase}\right].
\end{equation} 
Under coherent illumination by a plane wave of unit amplitude, the sample diffracts and produces, at distance $z$, a hologram $\V \hologram_z$ captured by the camera:
\begin{equation}\label{eq:imageFormationModel}
    \hologram_z=\model_z(\V{\transmittance})+\V{\noise} \textmd{\ \ with }  \model_z(\V{\transmittance})=\lvert \Hz \V{\transmittance} \rvert^2,
\end{equation}
where $\model_z()$ models the noise-free intensity of the diffraction pattern (i.e., the hologram formation model) and $\V{\noise}$ represents the measurement noise.
$\Hz \in \mathbb{C}^{N\times N}$, is a 2D convolution operator which models propagation of the complex amplitude of the wave exiting the object plane. The computation of the squared modulus at each pixel, $\lvert \cdot \rvert^2$,  models intensity-only recording.

In the following, we consider the angular spectrum~\cite{goodman2005introduction} to model forth-and-back-propagation.
The introduction of lenses between the sample and the camera produces both a magnification and, due to the finite size of the pupil, an optical cutoff frequency (i.e., a low-pass filtering effect).
The magnification can be accounted for by adjusting the pixel size.
In the following, $\Hz$ accounts for both free-space propagation and low-pass filtering introduced by optical elements.

In a simple linear reconstruction approach, the hologram is back-propagated directly to the sample plane as if the sensor had recorded the complex amplitude of the wavefront.
This method is computationally inexpensive, requiring a single convolution.
However, as the sensor only records the intensity of the wave, this approach results in the superimposition of an in-focus image of the object and an out-of-focus version of it, the so-called twin-image (See Appendix \ref{sec:Appendix_TwinImage}).
Losing the phase information in the sensor plane introduces morphological and quantitativity errors in the reconstructed plane.

\subsection{Inference step of the proposed approach}
To facilitate the network reconstruction task, it is common in inverse problems applications to feed the network with an approximate inversion of the image and add a skip connection to networks such that they only learn to recover the difference between the ideal inversion and the approximated one~\cite{mccann2017convolutional}.
In the case of linear inverse problems, this approximate inversion usually corresponds to the adjoint of the forward operator, or to its pseudo inverse.
In holography, it is common to feed the network with a back-propagation of the hologram~\cite{rivenson2018phase,huang2021holographic-3a2}.
This pre-processing of the data is also applied to our approach.
Our proposed network, $f_{\boldsymbol{\vartheta}}$ (parameterized by weights $\boldsymbol{\vartheta}$), reconstructs the sample's transmission function directly from the back-propagation of a single hologram ($\Hbz \V{\hologram_z}$).
\begin{equation}
\widehat{\V \transmittance}^{\textmd{OURS}}= f_{\widehat{\V \vartheta}}\left(\Hbz \V{\hologram_z} \right)
\label{eq:inference}
\end{equation}
With this back-propagation as a pre-processing of the data, the reconstruction task is simplified as it becomes an unmixing task that is directly performed in the object plane and consists in separating the contribution of the real image from those of the twin- and 0-order images (Appendix \ref{sec:Appendix_TwinImage}, eq.\ref{eq:TwinImage}).

\subsection{Physics and phase-diversity -based training}

The training step uses the phase diversity from two out-of-focus intensity measurements $\hologram_{z}$ and $\hologram_{z+\delta z}$ captured at the distances $z$ and $z+\delta z$.
To train the network to suppress the twin-image from the back-propagation, phase diversity is exploited by introducing two data-fidelity terms for the two holograms in the physics-based loss function $L_{z}$.
To avoid the optimization process to fall into a non-physical local minimum of the loss function, a regularization term applied on the reconstruction is added. 
Its weight $\alpha_\downarrow$ is decayed over the training.
Once the parameters of the network have converged to a well-behaving configuration, this hyperparameter is set to 0.
The resulting loss function is defined by:
\begin{equation}
\begin{split}
L(\boldsymbol \vartheta,(\hologram_z,\hologram_{z+\delta z}))&=\lVert \hologram_{z} - \model_z\left(f_{\boldsymbol \vartheta}\left(\Hbz \V{\hologram_z} \right)\right)  \rVert^2_2 \\
&+ \lVert \hologram_{z+\delta_z} - \model_{z+\delta z} \left( f_{\boldsymbol \vartheta}\left(\Hbz \V{\hologram_z} \right) \right)\rVert^2_2 \\
&+ \alpha_{\downarrow}\mathscr{R}\left( f_{\boldsymbol \vartheta}\left(\Hbz \V{\hologram_z} \right) \right).
\end{split}
\label{eq:loss}
\end{equation}

The reconstruction problem can be solved during the training step such that $f_{\boldsymbol \vartheta}(\Hbz \V{\hologram_z})$ provides an accurate transmission function, regarding the loss function, for all the hologram pairs $\{(\hologram_{z}$,$\hologram_{z+\delta z})_i\}$ of the training dataset:
\begin{equation}\label{eq:LossTrain}
\widehat{\V \vartheta} = \underset{\V \vartheta}{\operatorname*{arg\,min}} \ \sum_i L(\boldsymbol \vartheta,(\hologram_z,\hologram_{z+\delta z})_i)
\end{equation}
where $L(\boldsymbol \vartheta,(\hologram_z,\hologram_{z+\delta z})_i)$ stands for the loss function term described in Eq. \ref{eq:loss} for the $i^\textmd{th}$ pair of holograms of the batch.
In practice, to prevent border issues arising from the propagation, the loss is evaluated only in the center of each patch, dividing the image size by 2 in each dimension.
\subsection{Network architecture}\label{sec:networkArchitecture}
We selected a U-Net architecture~\cite{ronneberger2015u} with a residual connection~\cite{Dalsasso2021sar2sar}.
The residual connection, added to the popular U-Net architecture, allows simplifying the reconstruction task as the hologram back-propagation, provided as input to the network, is already an approximation of the final reconstruction (see section~\ref{sec:ImageFormationModel}).
In this framework, the network untangles the twin and real image contributions from the back-propagated images.

\section{Evaluation of the method on simulated data}
\label{sec:Recons_holo_simu}
In this section, our approach is validated using two simulated hologram datasets: one modeling rod-shaped bacteria (bacilli), and the other consisting of spherical particles (simulating calibration beads).
Our motivation for considering simulated holograms is to run well-controlled numerical experiments where ground-truth transmission functions $\boldsymbol t^{\textmd{GT}}$ can be used to conduct a quantitative evaluation of the reconstructed transmission functions.
These simulations are also used to perform a supervised training of a reference network to evaluate how closely our self-supervised approach can get, without ever using the ground-truth images during training.
First, we describe the simulation of the dataset; second, we discuss the metrics used to evaluate the performances of our algorithm; finally, we compare our approach with other state-of-the-art reconstruction methods.

\subsection{Simulation of hologram datasets}
\label{subsec:simulationholo}

Simulation parameters were chosen close to the experimental parameters (Sec. \ref{sec:Recons_holo_expe}). The wavelength, magnification, refractive indices and sensor parameters are given in Table.\ref{tab:experimental_parameters}.

\subsubsection{Simulated holograms of bacteria}
\label{subsubsec:SimHoloBacteria}
To evaluate the performance of the networks on simulated objects close to biological experimental ones, we generated a dataset of simulated bacteria holograms, referred in the following as ``SIM\_BACT''.
It contains 9000 paired samples (with train:validation:test ratios of 91\%:6\%:3\%).
To represent small bacteria in silicon oil, we simulated 0.5 µm thick 2D Béziers curves with a length drawn randomly from a uniform distribution in the range [1-2 µm], and sampled it on the 2D grid of the sensor.
The refractive index of the bacteria was chosen to match the experimental values.
The obtained transmission function, which is used as ground truth, is propagated to the sensor plane.
The hologram is then obtained by retaining the square modulus of this diffracted wave.
A white Gaussian noise is added, leading to a signal-to-noise ratio of 62.5. Figure~\ref{fig:bacteria_simulation_results} shows examples of paired simulated bacteria holograms and their corresponding ground truth transmission function.

\subsubsection{Simulation of bead holograms}
\label{subsubsec:SimHoloBeads}
Datasets of 9000 paired holograms (with train:validation:test ratios of 91\%:6\%:3\%), referred as ``SIM\_BEADS'' in the following, were simulated.
The simulated objects, latex beads in silicon oil, have all the same defocus distance.
Considering a coherent illumination by a uniform plane wave of unitary amplitude, the complex amplitude of the diffracted wave on the sensor plane is computed using the Lorenz–Mie model~\cite{slimani1984near}.
The hologram is obtained by taking the squared modulus of this complex amplitude.
The propagation distances are : $z=10$ µm ± $0.5$ µm and $\delta z= 4$ µm.
A white Gaussian noise $\V \noise$ is added to the simulated holograms, leading to a signal-to-noise ratio of 31.25 ($\text{SNR}=\left<\V \hologram_z\right>/\sigma$ where $\left<\V \hologram_z\right>$ is the spatial average of the hologram $\hologram_z$, and $\sigma$ the standard deviation of the noise), which matches the SNR of experimental data.
In these simulations, two populations of beads are simulated. In the following they are referred as main and secondary beads.
The main population corresponds to isolated objects around which the secondary bead's population makes clusters.
The positions and parameters of the beads are drawn from uniform laws while ensuring there is no overlap between the beads. 

Table~\ref{tab:beads_parameters} indicate the values of all the parameters used.
The ground truth transmission function, $\V \transmittance^{\textmd{GT}}$, is obtained by back-propagating the complex amplitude in the sensor plane to the object plane with an angular spectrum propagation~\cite{goodman2005introduction}.
This simulation dataset provides a much more complex example of samples with a higher diversity of objects than the SIM\_BACT and the experimental ones and aims at evaluating the robustness of our approach.
\begin{table}[t]
\caption{Table of parameters for the holograms simulation and experimental recording}
\label{tab:experimental_parameters}
\centering
\begin{tabular}{ m{.5\columnwidth} c }
  \toprule
  \bfseries Parameter & \bfseries Value\\
  \midrule
  Wavelength & 637.6 nm \\
  Input hologram size (in pixels) & 512x512 \\ 
  Pixel pitch & 11 µm \\
  Magnification & 67.5 \\
  Object space pixel pitch & 163 nm \\
  Refractive index of the immersion medium (silicon oil) & 1.402\\
  Numerical aperture & 1.3\\
    $z$ and  $\delta z$ & 10 µm ± 0.5 µm and 4 µm\\
  \bottomrule
\end{tabular}
\end{table}

\begin{table}[t]
\caption{Table of beads parameters for simulations}
\label{tab:beads_parameters}
\centering
\begin{tabular}{ m{.5\columnwidth} c }
  \toprule
  \bfseries Parameter & \bfseries Value\\
  \midrule
  Range of diameter for main beads & [0.5 µm, 5 µm] \\
  Range of diameter for secondary beads & [0.5 µm, 2.75 µm] \\
  Range of refractive index & [1.402, 1.713] \\
  \bottomrule
\end{tabular}
\end{table}
\subsection{Evaluation metrics of the reconstructions}\label{subsec:EvaluationMetrics}
The performance of the reconstruction algorithms on the simulated datasets is evaluated by comparing the phase and modulus of the reconstructions $\widehat{\V \transmittance}$, to those of the ground truth $\V \transmittance^{\textmd{GT}}$.
They were assessed using three metrics: the Root Mean Square Error (RMSE) over the entire field-of-view, the Objects Root Mean Square Error (RMSE-O), and the processing time:
\begin{itemize}
    \item The RMSE on the reconstructed modulus/phase of the transmission function provides a quantitative indicator of the accuracy of the reconstructions.
    \begin{equation}
    \text{RMSE}( \V u^{\textmd{GT}}, \widehat{\V u})=\sqrt{\frac{1}{N}\underset{k=1}{\overset{N}{\sum}}~|u^{\textmd{GT}}_k - \widehat{u}_k|^2}
    \end{equation}
    where $\V u$ can stand for the modulus or the phase of the transmission function.
    \item The RMSE-O accounts only for the objects of interest, without considering all the background of the transmission function (due to the sparsity of the samples, most of the pixels correspond to the background).

    \begin{equation}
        \text{RMSE-O}(\V \transmittance^{\textmd{GT}}, \widehat{\V \transmittance})=\sqrt{\frac{1}{M}\underset{k=1}{\overset{M}{\sum}}~w_k|\transmittance^{\textmd{GT}}_k - \widehat{\transmittance}_k|^2}
    \end{equation}
    with $M=\underset{k=1}{\overset{N}{\sum}}w_k$ and $w_k$ a binary mask at pixel position $k$, with value 1 granted if the pixel belongs to an object and 0 if it doesn't.
\end{itemize}

The metrics are presented as an average over the test image dataset.

\subsection{State-of-the-art methods used in the comparisons}\label{subsec:State-of-the-art methods}
Our method is compared to five state-of-the-art reconstruction approaches.
On the one hand, back-propagation, regularized IPA using one hologram (IPA-1) and supervised learning are methods that provide reconstructions by processing a single hologram.
On the other hand, both GS and IPA with two holograms (IPA-2) exploit phase diversity to achieve quantitative reconstructions, but they require two holograms to perform a reconstruction.
IPA-1 and IPA-2 reconstructions consist of minimizing Eq. \ref{eq:IPA} using a single (resp. two) hologram.
\begin{equation}\label{eq:IPA}
    \widehat{\V \transmittance}^{\textmd{IPA}}=\underset{\V \transmittance \in \mathbb{C}^N}{\operatorname*{arg\,min}}~\Bigl[\mathscr{D}(\V \transmittance)+\alpha \mathscr{R}(\V \transmittance)\Bigr]
\end{equation}
where
\begin{equation*}
\left\{
\begin{array}{l}
    \mathscr{D}(\V \transmittance) = \lVert \hologram_z -\model_z(\V \transmittance) \rVert_2^2 \ \textmd{(IPA-1)}\\
    \mathscr{D}(\V \transmittance) = \lVert \hologram_z -\model_z(\V \transmittance) \rVert_2^2 + \lVert \hologram_{z+\delta z} -\model_{z+\delta z}(\V \transmittance) \rVert_2^2 \ \textmd{(IPA-2)}
\end{array}
\right.
\end{equation*}
The noise is assumed to be Gaussian, therefore, the L$^2$ norm, $\lVert \cdot \rVert_2$, is used to compute the data-fidelity terms $\mathscr{D}$.
For regularization, we chose a piecewise constant prior, expressed as a hyperbolic total variation term:
\begin{equation}
    \mathscr{R}(\V \transmittance)=\frac{1}{N}\underset{k=1}{\overset{N}{\sum}}~\sqrt{|\nabla_{x, k}\V \transmittance|^2+|\nabla_{y, k} \V \transmittance|^2+\epsilon}
\end{equation}
where the gradient operators $\nabla_{x, k}, \nabla_{y, k}$ corresponds to the finite differences at pixel $k$ in the $x$ and $y$ directions, and $\epsilon$ is a hyperparameter to make the TV term differentiable near 0.
In simulation, $\alpha$ is tuned optimally regarding the minimum of RMSE on the phase reconstruction and $\epsilon$ is set manually to the value $10^{-6}$. 
In experimental conditions, both regularization hyperparameters are tuned manually to the values of $\alpha = 0.05$ and $\epsilon=10^{-6}$. Table~\ref{tab:hyperprameters} summarizes the values of the hyperparameters that where used for IPA reconstructions.

\begin{table}[t]
    \caption{Regularization hyperparameters for IPA approaches}
    \label{tab:hyperprameters}
    \centering
    \begin{tabular}{ m{.2\columnwidth} c c c}
        \toprule
        \bfseries Dataset & $\alpha_{\textmd{IPA-1}}$ & $\alpha_{\textmd{IPA-2}}$ & $\epsilon$\\
        \midrule
        SIM\_BACT & 0.05 & 0.1 & 1e-6 \\
        SIM\_BEADS & 0.1 & 0.025 & 1e-6\\
        EXP\_BEADS & 0.05 & 0.05 & 1e-6 \\
        EXP\_ECOLI & 0.05 & 0.05 & 1e-6 \\
        EXP\_MLUTEUS & 0.05 & 0.05 & 1e-6 \\
        \bottomrule
    \end{tabular}
\end{table}

\subsection{Training on simulated data}
\label{subsec:TrainingSimuData}

The neural network is trained over 500 epochs, with a learning rate of 0.002, a batch size of 8 and a dropout regularization of 10\%.
The learning rate is updated by a scheduler that reduces it on plateau.
Parameters $\boldsymbol \vartheta$ are then optimized using the AMSGRAD optimizer~\cite{reddi2018adam-and-beyond}.
The TV regularization hyperparameter $\alpha_\downarrow$ of the training loss (eq. \ref{eq:loss}) is updated at each epoch $e$ with the formula $\alpha_{\downarrow}^{(e)}=\text{max}(0, 0.5-0.002e)$.
Therefore, the regularization only serves as a guide for the training's first steps and is completely discarded at half-training time, allowing the network to free itself from this handcrafted prior. 
Reconstruction on an EXP\_BEADS image without regularization is shown in the supplementary materials Fig. 7. It shows low-frequencies oscillations artifacts that illustrates the need of the regularization term at the beginning of the training.
Our approach is implemented with Pytorch, and the code is available at https://gitlab.univ-st-etienne.fr/rif02493/self-supervised-dihm.
To validate its performance, we performed the experiments with a Graphics Processing Unit NVIDIA RTX A6000.
\begin{table}[t]
    \caption{Training parameters}
    \label{tab:params_training}
    \centering
    \begin{tabular}{ m{.5\columnwidth} c }
        \toprule
        \bfseries Parameter & \bfseries Value\\
        \midrule
        Epochs & 500 \\
        Learning rate & 0.002 \\
        Batch size & 8 \\
        Dropout percentage & 10 \\
        Optimiser & AMSGRAD \\
        \bottomrule
    \end{tabular}
\end{table}

\subsection{Reconstruction of simulated data}
\label{subsec:ReconsSimuData}
\begin{figure*}[t]
    \centering
    \includegraphics[width=\linewidth]{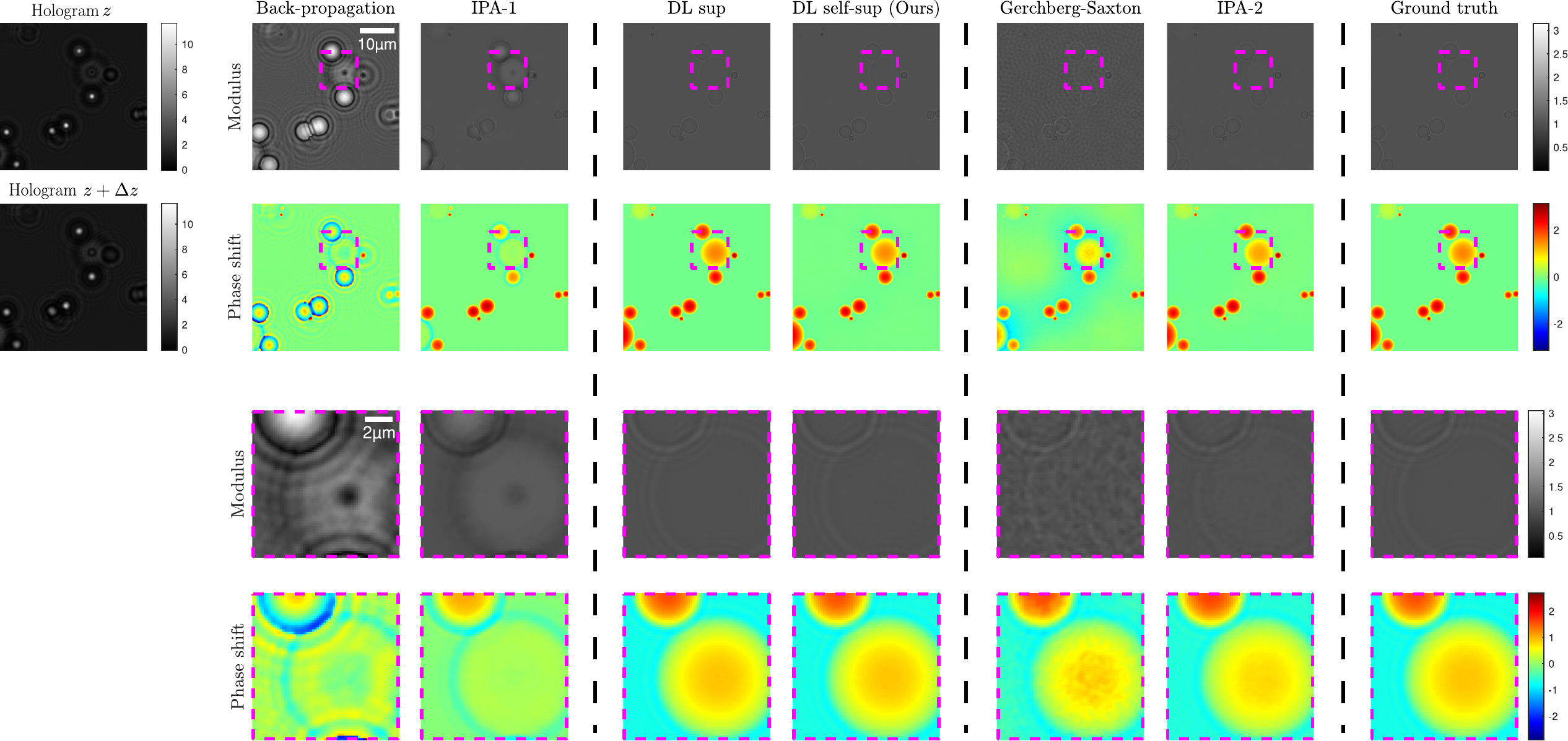}
    \caption{Reconstruction of a test image of the the SIM\_BEADS dataset using state-of-the-art approaches and our proposed method. Deep learning reconstructions are trained on the SIM\_BEADS dataset. The purple squares represent the zoom region, of size 64×64 pixels.} 
    \label{fig:bacteria_simulation_results}
\end{figure*}

First, we provide a qualitative analysis of the reconstruction results on the SIM\_BEADS dataset.
Figure~\ref{fig:bacteria_simulation_results} shows an example of phase reconstructions using five state-of-the-art methods and our proposed approach, on a simulated hologram of the SIM\_BEADS test dataset. A similar example on the SIM\_BACT test dataset is provided in the supplementary materials Fig. 5.
As can be seen in Fig.~\ref{fig:bacteria_simulation_results}, the back-propagation reconstruction suffers from twin-image artifacts that may lead to segmentation errors and/or misinterpretation of both the reconstructions in modulus and in phase.
The GS algorithm provides a more accurate reconstruction with reduced twin-image artifacts.
This confirms that exploiting phase diversity allows solving the ill-posed phase retrieval problem in DIH. However, as there is no other constraint on the reconstructed image, the algorithm is not robust to noise.
Indeed, the GS algorithm fits the noise, leading to noisy reconstructions.
In this example, regularized IPA with a single hologram fails to produce an accurate quantitative reconstruction on all the objects.
This is due to the high diversity of objects in size and phase shift in this dataset.
Indeed, even though the weight of the total variation is tuned optimally, this regularization prior is not adapted for all objects.
In comparison, our method provides better results, close to supervised deep learning and regularized IPA with two holograms while being self-supervised and requiring a single hologram for inference.

This qualitative analysis is confirmed by the performance metrics values that are provided in Table \ref{tab:res_sim_phase} evaluated on the phase of 300 reconstructed holograms from both SIM\_BEAD and SIM\_BACT validation dataset.
Indeed, the back-propagation and GS methods provide the worst RMSE and RMSE-O values, while our approach and supervised deep learning methods provide the best.
In these simulations, supervised deep learning represents the highest performance that is achievable by the network architecture because it is directly trained on ground truth data.
However, in real cases this ground truth is hardly accessible.
With the same number of parameters, our self-supervised approach achieves similar performances.
The RMSE/RMSE-O are of the same order of magnitude with values of only of 4.75/12.9 mrad on the SIM\_BEADS dataset and 4.61/15.2 mrad on the SIM\_BACT dataset.
With this accuracy, our self-supervised approach is significantly more accurate than the GS ones.
Comparison with regularized IPA emphases the ability of our approach to reconstruct the samples.
Indeed, while on SIM\_BACT, both RMSE and RMSE-O for IPA-1, IPA-2 and our approach are close, on the more complex SIM\_BEADS dataset, our approach performs significantly better. The RMSE and RMSE-O are almost divided by 2 in comparison with IPA-2, which requires two holograms for reconstruction, and by 10 compared to IPA-1.
The inference computational time of our approach, is also much lower compared to state-of-the-art reconstruction methods, and improved even further when reconstructing multiple images simultaneously (batched inference).
Table~\ref{tab:res_sim_phase} provides the computational time of all reconstruction methods on 300 simulated images, showing that our approach achieves similar or better performances than IPA methods while being 1000 times faster.

To investigate our network's ability to reconstruct objects outside the distribution of the training dataset, we evaluated its performance when trained and evaluated on two different datasets (see Table \ref{tab:res_sim_phase}). In this case, the network does not improve the back-propagation reconstruction. 
Indeed, as the network learns an implicit prior on the distribution of the target objects, it is advised to train it on a specific dataset of interest.

In the appendix~\ref{sec:Appendix_Modulus_rec}, we provide the RMSE and RMSE-O values for the modulus.
The analysis of these metrics leads to similar conclusions.

\begin{table*}[t]
    \caption{Reconstruction metrics on the phase of simulated datasets over 300 test images}
    \label{tab:res_sim_phase}
    \centering
    \begin{tabular}{ m{0.5\columnwidth} c c c c c c c c }
        \toprule
        \bfseries{Method} & \makecell{\bfseries{Training}\\ \bfseries{set}}  & \multicolumn{5}{c}{\bfseries{Test dataset}}& \makecell{\bfseries{\# of holograms used}\\ \bfseries{for reconstruction}} & \makecell{\bfseries{Reconstruction}\\ \bfseries{time}}\\
        \midrule
        & & \multicolumn{2}{c}{SIM\_BACT} & & \multicolumn{2}{c}{SIM\_BEADS} & &\\
        \midrule
             & & \makecell{RMSE\\(mrad)} & \makecell{RMSE-O\\(mrad)}  & & \makecell{RMSE\\(mrad)} & \makecell{RMSE-O\\(mrad)}  \\
            \cmidrule{3-4} \cmidrule{6-7}
            Back-propagation & & 42.5 & 138 & & 406 & 1180 & 1 & \textbf{0.1 ms}\\
            IPA-1 & & 4.68 & 27.0 & & 245 & 750  & 1 & 5.6 s\\
            Gerchberg-Saxton & & 35.3 & 54.7  & & 176 & 387  & 2 & 4.6 s\\
            IPA-2 & & \underline{3.41} & \textbf{19.3}  & & 41.1 & 100  & 2 & 13.3 s\\
            \midrule 
            Supervised DL & \multirow{2}{*}{SIM\_BACT} & \textbf{2.92} & \underline{23.3}   & & 270 & 778  & 1 & \underline{8.1/1.1 ms*}\\
            Self-supervised DL (ours) & & 7.56 & 24.1  & & 415 & 1280  & 1 & \underline{8.1/1.1 ms*} \\
            \midrule
            Supervised DL & \multirow{2}{*}{SIM\_BEADS} & 11.1 & 98.2 & & \textbf{4.75} & \textbf{12.9}  & 1 & \underline{8.1/1.1 ms*}\\
            Self-supervised DL (ours) & & 15.9 & 136  & & \underline{22.7} & \underline{56.3}  & 1 & \underline{8.1/1.1 ms*}\\
        \bottomrule
        *unbatched/batched (16 images)
    \end{tabular}
    
\end{table*}

\section{Reconstruction of experimental holograms}
\label{sec:Recons_holo_expe}
To evaluate our method's performance in experimental conditions, we recorded three experimental hologram datasets of monodispersed calibration beads and bacteria.
We compare the performance of our approach when applied to these experimental datasets with that of the state-of-the-art methods.

\subsection{Experiments}

\begin{figure*}[t]
    \centering
    \includegraphics[width=0.75\linewidth]{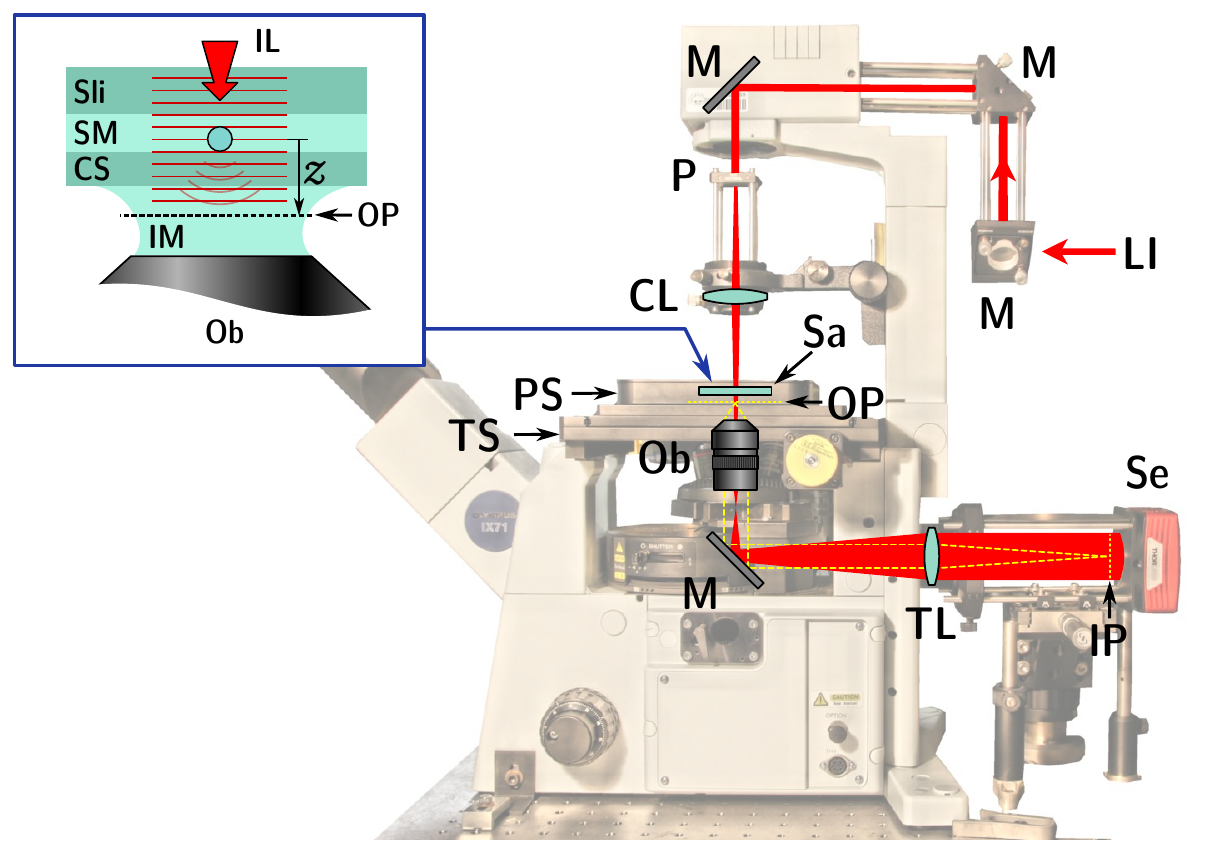}
    \caption{\textbf{Experimental setup:} LI: Laser Illumination, M: Mirrors, P: Pinhole, CL: Condenser Lens, Sa: Sample, PS: Piezo Stage, TS: Translation Stage, OP: Object Plane, Ob: Objective, TL: Tube Lens, IP: Image Plane, Se: Sensor. \textbf{Inset:} details of the sample. IL: Illumination Light, Sli: Slide, SM: Sample Medium (Silicon oil and micro-objects), CS: Coverslip, OP: Object Plane, IM: Immersion Medium (Silicon oil), Ob: Objective, $z$: defocus distance (distance from the micro-object to the object plane, defined as the plane conjugated with the sensor plane) }
    \label{fig:setup}
\end{figure*}

Images of the datasets were acquired using an Olympus inverted microscope adapted for in-line holography (see Figure~\ref{fig:setup}).
The experimental parameters are the same  as the ones used for the simulations (Table \ref{tab:experimental_parameters}).
The illumination source is both spatially and temporally coherent and can be considered a plane wave.
The image datasets were acquired using a silicon oil immersion objective with a numerical aperture of 1.3 and a tube lens in a telecentric configuration.
Using a microscopy target, the magnification has been estimated to be 67.5.
The equivalent pixel size in the object plane is \SI{163}{nm}.

Three different samples were prepared: polystyrene calibration beads (``EXP\_BEADS'') and two types of bacteria: a Gram-negative bacillus, \emph{Escherichia Coli} (``EXP\_ECOLI''), and a Gram-positive coccus, \emph{Micrococcus Luteus} (``EXP\_MLUTEUS'').
They were spread on microscope coverslips in silicon oil.
A more detailed version of the setup description, sample preparation and acquisition can be found in the supplementary materials~\ref{sec:Appendix_ExpeDetails}.

Images acquired with our experimental setup are cropped into sub-images of size 512$\times$512 pixels, and are split into training and validation sets in a (95\%,2.5\%,2.5\%) ratio for train, validation and test respectively.
This yields 7168 images for EXP\_BEAD, 7003 images for EXP\_ECOLI, and 10781 images for EXP\_MLUTEUS.
These image have been distributed with train:validation:test ratios of 90\%,5\%,5\% for each experimental dataset. All experimental datasets are available at https://labh-curien.univ-st-etienne.fr/public-data/BioDIHM/.
The neural network is trained on each training dataset using the same hyperparameters as those used for the simulation tests.
Training on images of the bead samples uses a regularization hyperparameter update rule defined by $\max(0.05, 0.5-0.002e)$ with $e$ the epoch number.

\subsection{Validation of our approach on experimental data}

For experimental data, the ground truth is not known; thus, we compared the reconstruction to a baseline computed with a TV-regularized inverse problems approach that uses two holograms (IPA-2) to perform the reconstruction. As seen in the simulations, such an approach gives reconstructions close to the ground truth.
The method's performance on the experimental data is evaluated in a similar way to that previously done on the simulated data (see Sec.\ref{subsec:EvaluationMetrics}), except that the ground truth is replaced by the baseline reconstruction  $ \widehat{\V \transmittance}^{\text{IPA-2}}$.

Prior to reconstruction, we estimated the defocus distances of the hologram using the commonly used GRA sharpness criterion~\cite{langehanenberg2011autofocusing}.
This criterion involves maximizing the gradient of the back-propagation with respect to the distance $z$ to retrieve the in-focus plane.
For real-world data, holograms at the distance $z$ and $z+\delta z$ often suffer from a slight shift in the $x$ and $y$ directions due to misalignment of the optical and mechanical components of the experimental setup.
Our algorithm accounts for these shifts, computed using the correlation between the two back-propagated holograms, in the forward model. In our experiment the average shift is about one-third of a pixel (\SI{54}{nm}) along the horizontal axis and negligible on the vertical axis.

\begin{figure*}[!t]
    \centering
    \includegraphics[width=0.9\linewidth]{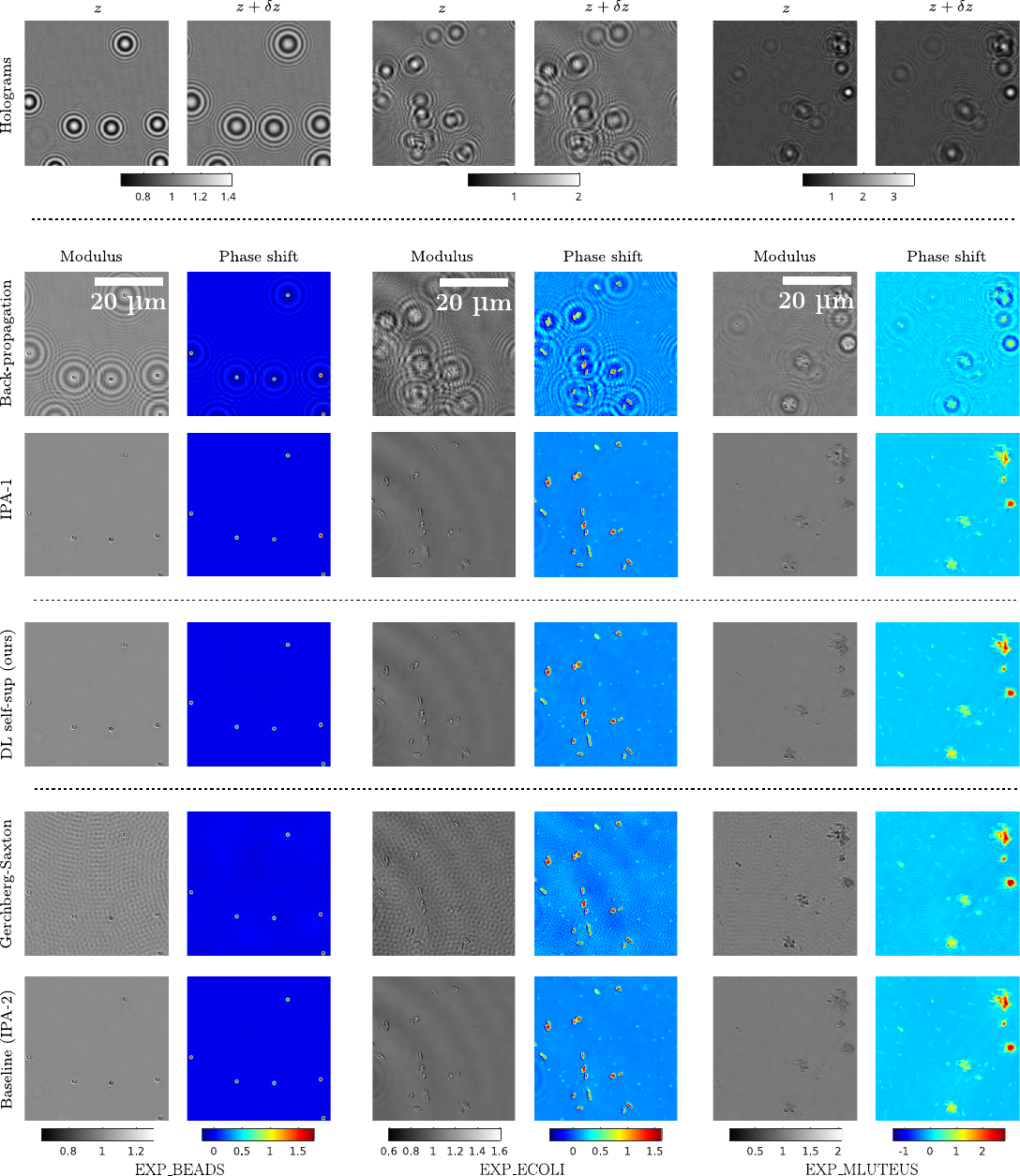}
    \caption{Comparison between state-of-the-art reconstructions methods and ours on three different samples types from the experimental datasets. Our approach produces significantly less artifacts than most reconstruction methods, including Gerchberg-Saxton which uses two holograms for reconstruction.}
    \label{fig:res_expe:visu_bact_e_coli}
\end{figure*}

Figure~\ref{fig:res_expe:visu_bact_e_coli} shows the result of the state-of-the-art methods (\ref{subsec:State-of-the-art methods}) and the proposed method on the EXP\_BEADS, EXP\_ECOLI and EXP\_MLUTEUS datasets. Zooms on the reconstructions are provided in supplementary material Fig.6.
As observed for simulation data, the back-propagation and GS approaches provide unsatisfactory reconstructions, which are either disturbed by twin-image or noise contamination.
Our proposed method provides a more satisfying reconstruction that is free of twin-image artifacts and close to the ground truth.
Visually, these reconstructions are similar to reconstructions provided by IPA-1.
The RMSE and RMSE-O values in Table~\ref{tab:res_expe:phase_comparison} support this qualitative analysis.
Quantitatively, our method achieves lower RMSE and RMSE-O than IPA-1 on all the experimental datasets.
Among all the compared methods, ours is the fastest one with a computational reconstruction time of a few milliseconds.
This shows the potential of our network to perform accurate live imaging reconstructions with a single hologram.

\begin{table*}[t]
    \caption{Comparison of phase reconstruction metrics on experimental test datasets}
    \label{tab:res_expe:phase_comparison}
    \centering
    \begin{tabular}{ m{0.4\columnwidth}  c c   c  c c   c  c c  c}
        \toprule
        \bfseries{Methods} & \multicolumn{8}{c}{\bfseries{Test dastaset}} & \makecell{\bfseries{\# of holograms used}\\ \bfseries{for reconstruction}}\\
        \midrule
        & \multicolumn{2}{c}{EXP\_BEADS} & & \multicolumn{2}{c}{EXP\_ECOLI} & & \multicolumn{2}{c}{EXP\_MLUTEUS}\\
        \midrule
        & \makecell{RMSE\\(mrad)} & \makecell{RMSE-O\\(mrad)}  & & \makecell{RMSE\\(mrad)} & \makecell{RMSE-O\\(mrad)}  & & \makecell{RMSE\\(mrad)} & \makecell{RMSE-O\\(mrad)} \\
        Back-propagation & 34.3 & 198   & & 92.2 & 313  & & 108 & 408 & 1\\
        IPA-1 & \underline{5.98} & \underline{75.9}  & & 23.4 & 92.6  & & 39.1 & 163 & 1\\
        Gerchberg-Saxton & 18.4 & 86.6 & & 44.5 & 101 & & 51.4 & 135 & 2\\
        \midrule
        Self-supervised DL (Ours) & \textbf{5.10} & \textbf{63.6}  & & \textbf{22.0} & \textbf{82.3}  & & \textbf{31.5} & \textbf{112} & 1\\
        
        \bottomrule
    \end{tabular}
    
\end{table*}

\section{Conclusion}
In this paper, we proposed a self-supervised physics-based deep learning approach for reconstructing the transmission function of a sample from a single digital in-line hologram.
Our approach exploits phase diversity during training by means of two defocused measurements.
The loss function to train the network relies on an accurate modeling of image formation that is used as a physical constraint to train the reconstruction network.
After training, the network can be used for fast inference by applying it to a simple backpropagation of the input hologram.
The performance of this method has been compared to other phase retrieval techniques.
On simulated datasets, the method provides accurate phase reconstructions, comparable to regularized inverse problems approaches and supervised learning in simple cases and outperforming regularized inverse problem approaches in more complex cases.
In contrast to regularized inverse problems approaches, our approach does not rely on priors in the object space or delicate hyperparameter tuning.
The computational time for inference is also a major advantage of our method, with a speedup by three orders of magnitude, which makes live reconstruction of biological samples possible.
Regarding experimental holograms, the method yields phase reconstructions of similar quality to IPA-1 regarding the RMSE and the RMSE on the objects.
In addition to the proposed method, we provide two simulated datasets or beads and bacteria, and three experimental datasets of beads, E. Coli and M-Luteus bacteria acquired with a DIHM setup available at https://labh-curien.univ-st-etienne.fr/public-data/BioDIHM/.
To ensure the reproducibility of the results, the code of our deep neural networks is made available at https://gitlab.univ-st-etienne.fr/rif02493/self-supervised-dihm.

\section*{Acknowledgments}
The authors would like to thank Jean-Louis Magnard (Laboratoire BVpam, UMR CNRS 5079, University Jean Monnet, Saint-Etienne) and technicians of the biology department of the faculty of Sciences and Techniques of University Jean Monnet, Saint-Etienne, for sharing their advice, competence and materials for the preparation of the bacteria samples.

{\appendices
\section{Twin image issue}
\label{sec:Appendix_TwinImage}
The back-propagation of the hologram leads to the reconstruction of the in-focus object transmission function mixed with unwanted signal. Indeed, by considering the difference $\deltaobjtransmcplx$ between the transmittance and a unitary plane wavefront (such that \V{\transmittance} = $\V{1} + \deltaobjtransmcplx $), the back-propagation of a hologram is modeled by:
\begin{equation*}
    \Hbz \V{\hologram_z}
    = \Hbz | \Hz (\V{1} +  \V{\deltaobjtransmcplx}) |^2\approx\Hbz | \V{1} + \Hz \V{\deltaobjtransmcplx} |^2
\end{equation*}
considering the propagation operator over a distance z, $\Hz$, modified not to consider the constant phase shift, $kz$, due to propagation along the z-axis such that $\Hz\V{1}\approx \V{1}$ (property of the operator $\Hz$\footnote{Physically, the propagation of a plane wave gives a plane wave. Due to the sensor field truncation, this is only approximately verified.}).
By developing the square modulus and considering that $\Hbz\Hz=\Hz^{\dagger}\Hz\approx\M I$ where $\Hz^{\dagger}$ is the Hermitian transpose\footnote{Physically, a propagation followed by a back-propagation does not change the wave amplitude~\cite{liebling2003fresnelets}. Due to the sensor field truncation and numerical aperture filtering, this operator is not the exact inverse of the propagation.}:
\begin{equation}\label{eq:TwinImage}
    \Hbz \V{\hologram_z}
    \approx \underbrace{\V{1}+
    \Hbz|\Hz\V{\deltaobjtransmcplx \mathstrut}|^2
    }_{\textmd{0 order}}
    + \underbrace{
    \V{\deltaobjtransmcplx \mathstrut}
    }_{ \textmd{+1 order}}
    + \underbrace{\Hbdz \V{\deltaobjtransmcplx^* \mathstrut}
    }_{\textmd{-1 order}}
\end{equation}
The different terms of this equation can be interpreted as follows:
\begin{itemize}
    \item a low-frequency 0-order term that contains: a term corresponding to the backpropagation of the intensity of the object wave propagated on the sensor (without interference);
\item a +1 order term corresponding to the object (called the real image);
\item a -1 order term corresponding to the conjugate of this object propagated by a distance $-2 z$.
    This contribution is that of a twin image (or virtual image) positioned symmetrically to the real image with respect to the hologram.
\end{itemize}

This last term is the most significant term that degrades the back propagation reconstruction. It is known as the twin image issue. Thus, reconstructing the sample's transmission function from the backpropagation of a hologram can be seen as an unmixing problem. 

\section{Experimental setup}
\label{sec:Appendix_ExpeDetails}

\subsubsection{Experiments}

Images of the datasets were acquired using an IX71 inverted microscope from Olympus that has been adapted for in-line holography (see Figure~\ref{fig:setup}). The illumination source is an Oxxius LBX-638 laser diode emitting at \SI{637.6}{nm} and coupled in a monomode fiber. The output of the fiber is then magnified with a microscope objective to illuminate a \SI{200}{\micro\meter}-diameter pinhole placed in the front focal plane of a lens. The focal length of the lens is \SI{100}{mm}. This lens replaces the usual condenser lens of the microscope, allowing for the setting of a spatially coherent illumination. The sample is placed near the back focal plane of the lens and is therefore illuminated by an Airy pattern that is large enough to cover the whole field of view without introducing stray light in the optical system. In these conditions, the illumination of the sample can be considered to be a plane wave. The inverted microscope is equipped with a Corvus stepper-motor XY-translation stage and a piezo Z-stage (P737.5SL from PI, with \SI{1.6}{nm} resolution and \SI{5}{nm} repeatability).

The image datasets used in this work were acquired with an Olympus UPlanSApo 60$\times$ silicon oil immersion objective with a numerical aperture of 1.3 and a coverslip thickness correction collar. The refractive index of silicon oil is 1.402 at \SI{637.6}{nm}. The silicon oil is also used as the sample medium. It is viscous enough to prevent any motion of the micro-objects during acquisitions and its relatively low refractive index enhances the contrast of the holograms. Moreover, silicon oil is not hygroscopic like glycerol, which guarantees a high stability of the refractive index. The optical system is then complemented with a \SI{200}{mm}-focal length apochromatic tube lens designed to work with infinity-corrected objectives in a telecentric configuration (TTL200MP from Thorlabs). In this configuration, the magnification has been calibrated to be 67.5 with a microscopy target. The images are finally acquired with a S805MU1 monochrome CCD sensor from Thorlabs. This sensor has 3296$\times$2472 pixels with a pixel size of \SI{5.5}{\micro\meter}. In these conditions, the field of view is 268.5$\times$201.4\si{\micro\meter}. Every image is downsampled by a factor 2, yielding a final pixel size of \SI{163}{nm} at the object plane.

\subsubsection{Samples preparation}

Two different types of microscopy samples have been tested here: transparent calibration beads and Gram-stained bacteria. Moreover, two types of non-pathogenic and very common bacteria have been tested: a Gram-negative bacillus (\emph{Escherichia Coli}) and a Gram-positive coccus (\emph{Micrococcus Luteus}). All the samples have been spread and prepared (see below) on a 22$\times$22\si{mm} coverslip (Precision cover glasses No.~1.5H) with a thickness of 0.17$\pm$0.005\si{mm}. 
Once completely dried, a droplet of silicon oil (Olympus silicon immersion oil for microscopy, SIL300CS-30CC) was poured on a standard glass microscopy slide (75$\times$26$\times$1\si{mm}) and the coverslip was flipped on it and sealed with varnish.
\begin{itemize}
\item \emph{Preparation of beads~:} The beads are NIST-certified polystyrene beads from ThermoFischer Scientific in aqueous suspension (10\%w/w) whose size is 1.0$\pm$0.03\si{\micro\meter} in diameter. Their refractive index at \SI{637.6}{nm} is 1.587. First a dilution of this suspension is done in ethanol with a ratio 1:1500 and then \SI{5}{\micro\liter} of this dilution was spread on the coverslip with an inoculation loop or the tip of the pipette.
\item \emph{Preparation of bacteria~:} both bacteria types were incubated at \SI{37}{\celsius} overnight to reach their stationary phase. Their growth was controlled with the standard optical density measurement at \SI{600}{nm}. Before slide preparation, the stationary culture is diluted in a liquid culture medium with a ratio of 1:20 and incubated again at \SI{37}{\celsius} until the bacteria population is doubled to reach an optical density approximately equal to 0.3. Then, \SI{3}{\micro\liter} of bacteria suspension is mixed to \SI{27}{\micro\liter} ethanol and spread on the coverslip with an inoculation loop to ensure bacteria fixation and homogeneous spreading. Then, the Gram-staining procedure consists in flooding the coverslip for \SI{1}{\minute} in Crystal Violet (primary stain), \SI{1}{\minute} in Lugol (mordant), \SI{5}{\second} in acetone/ethanol (decolorizer) and \SI{1}{\minute} in Safranin (counterstain). The sample is moreover rinsed with water between each step. After this procedure, as they are Gram-negative, \emph{E. Coli} are colored in pink, because the Crystal Violet dye was bleached by the acetone/ethanol mix. On the contrary, Gram-positive bacteria like \emph{M. Luteus} are colored in dark purple. Therefore, Gram-stained \emph{E. Coli} and \emph{M. Luteus} are supposed to be weakly absorbing at \SI{637.6}{nm}, with \emph{M. Luteus} absorbing a little more and are likely to have different dephasing properties~\cite{brault_multispectral_2023}.
\end{itemize}

\subsubsection{Aquisition and preprocessing}

The XY-stage, the Piezo Z-stage and the sensor are controlled with \textmu Manager~\cite{edelstein_computer_2010}. This allowed to perform automatized acquisition of contiguous, but non-overlapping fields of views of our different samples (typically 10$\times$10 to 16$\times$16 fields of view, depending on the sample density). For each field of view, 3 images are recorded, one in focus and 2 at different defocused positions (10 and \SI{14}{\micro\meter}, see subsection~\ref{subsec:simulationholo}). Because the slide is usually tilted relatively to the optical axis, the focus position can change, typically within $\pm\SI{3}{\micro\meter}$ when 16$\times$16 fields of view are acquired (corresponding to an area of 4.5$\times$3.5\si{mm} explored on the sample). To correct this, the best focus positions are evaluated in the four corners of the total explored area and corrected during the acquisition of the various fields of view. Because of its nanometric resolution and repeatability, the $\delta z$ movement of the piezo stage between the two defocused positions is known very precisely.

Additionally, a set of images (as a function of time) is recorded with the laser diode off to calculate an average black level image that is subtracted from every image. A background image is also estimated by calculating the median value of all the images at focus. All images are then normalized by dividing them by the background image.

\section{Supplementary results : modulus reconstruction}\label{sec:Appendix_Modulus_rec}
In this section we provide the performances of our proposed approach on modulus estimations for both simulated (Table \ref{tab:res_sim_mod}) and experimental datasets (Table \ref{tab:res_expe:modulus_comparison}). Analysis of these results lead to similar conclusions as the ones described for phase in the main manuscript.

\begin{table*}[t]
    \caption{Reconstruction metrics on the modulus of simulated datasets over 300 test images}
    \label{tab:res_sim_mod}
    \centering
    \begin{tabular}{ m{0.5\columnwidth} c c c c c c c c }
        \toprule
        \bfseries{Method} & \makecell{\bfseries{Training}\\ \bfseries{set}}  & \multicolumn{5}{c}{\bfseries{Test dataset}}& \bfseries{\# of holograms} & \makecell{\bfseries{Reconstruction}\\ \bfseries{time}}\\
        \midrule
        & & \multicolumn{2}{c}{SIM\_BACT} & & \multicolumn{2}{c}{SIM\_BEADS} & &\\
        \midrule
             & & \makecell{RMSE\\(×$10^{-3}$)} & \makecell{RMSE-O\\(×$10^{-3}$)}  & & \makecell{RMSE\\(×$10^{-3}$)} & \makecell{RMSE-O\\(×$10^{-3}$)}  \\
            \cmidrule{3-4} \cmidrule{6-7}
            Back-propagation & & 39.8 & 88.0 & & 200 & 556 & 1 & \textbf{0.1 ms}\\
            IPA-1 & & 3.29 & 13.1 & & 40.5 & 108  & 1 & 5.6 s\\
            Gerchberg-Saxton & & 34.7 & 45.7  & & 40.0 & 45.1  & 2 & 4.6 s\\
            IPA-2 & & \underline{2.44} & \textbf{10.8}  & & 9.50 & 14.9  & 2 & 13.3 s\\
            \midrule 
            Supervised DL & \multirow{2}{*}{SIM\_BACT} & \textbf{1.96} & \underline{13.7}   & & 108 & 299  & 1 & \underline{8.1/1.1 ms*}\\
            Self-supervised DL (ours) & & 5.11 & 14.9  & & 3450 & 4040  & 1 & \underline{8.1/1.1 ms*} \\
            \midrule
            Supervised DL & \multirow{2}{*}{SIM\_BEADS} & 6.44 & 66.3  & & \textbf{2.73} & \textbf{5.37}  & 1 & \underline{8.1/1.1 ms*}\\
            Self-supervised DL (ours) & & 8.11 & 61.1  & & \underline{5.90} & \underline{7.79}  & 1 & \underline{8.1/1.1 ms*}\\
        \bottomrule
        *unbatched/batched (16 images)
    \end{tabular}
    
\end{table*}

\begin{table*}[t]
    \caption{Comparison of modulus reconstruction metrics on experimental test datasets}
    \label{tab:res_expe:modulus_comparison}
    \centering
    \begin{tabular}{ m{0.4\columnwidth}  c c   c  c c   c  c c  c}
        \toprule
        \bfseries{Methods} & \multicolumn{8}{c}{\bfseries{Test dastaset}} & \bfseries{\# of holograms}\\
        \midrule
        & \multicolumn{2}{c}{EXP\_BEADS} & & \multicolumn{2}{c}{EXP\_ECOLI} & & \multicolumn{2}{c}{EXP\_MLUTEUS}\\
        \midrule
        & \makecell{RMSE\\(×$10^{-3}$)} & \makecell{RMSE-O\\(×$10^{-3}$)}  & & \makecell{RMSE\\(×$10^{-3}$)} & \makecell{RMSE-O\\(×$10^{-3}$)}  & & \makecell{RMSE\\(×$10^{-3}$)} & \makecell{RMSE-O\\(×$10^{-3}$)} \\
        Back-propagation & 31.0 & 103   & & 75.7 & 168  & & 70.5 & 190 & 1\\
        IPA-1 & \textbf{3.30} & \textbf{36.5}  & & \underline{10.2} & \underline{35.6}  & & \underline{12.7} & 43.8 & 1\\
        Gerchberg-Saxton & 18.7 & \underline{38.5} & & 37.8 & 42.4 & & 34.9 & \underline{41.8} & 2\\
        \midrule
        Self-supervised DL (Ours) & \underline{3.40} & 40.6  & & \textbf{8.52} & \textbf{32.3}  & & \textbf{10.6} & \textbf{32.1} & 1\\
        
        \bottomrule
    \end{tabular}
    
\end{table*}

}

\bibliographystyle{IEEEtran}
\bibliography{main}

\section{Biography Section}

\begin{IEEEbiography}[{\includegraphics[width=1in,height=1.25in,clip,keepaspectratio]{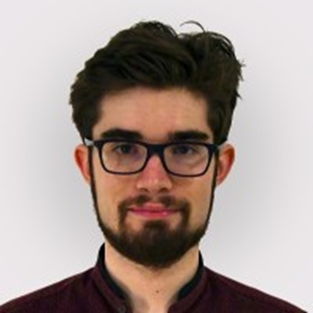}}]{Dylan Brault} received his Ph.D. degree in Image, Vision and signal science from Université Jean Monnet (France) in 2022. He is currently a postdoctoral fellow at the Paul Scherrer Institute (Switzerland).  His research interests include computational imaging, inverse problems, design of optical-digital imaging systems, physics informed machine learning and image and volume reconstruction with applications mostly related to biomedical imaging. 
\end{IEEEbiography}

\begin{IEEEbiography}[{\includegraphics[width=1in,height=1.25in,clip,keepaspectratio]{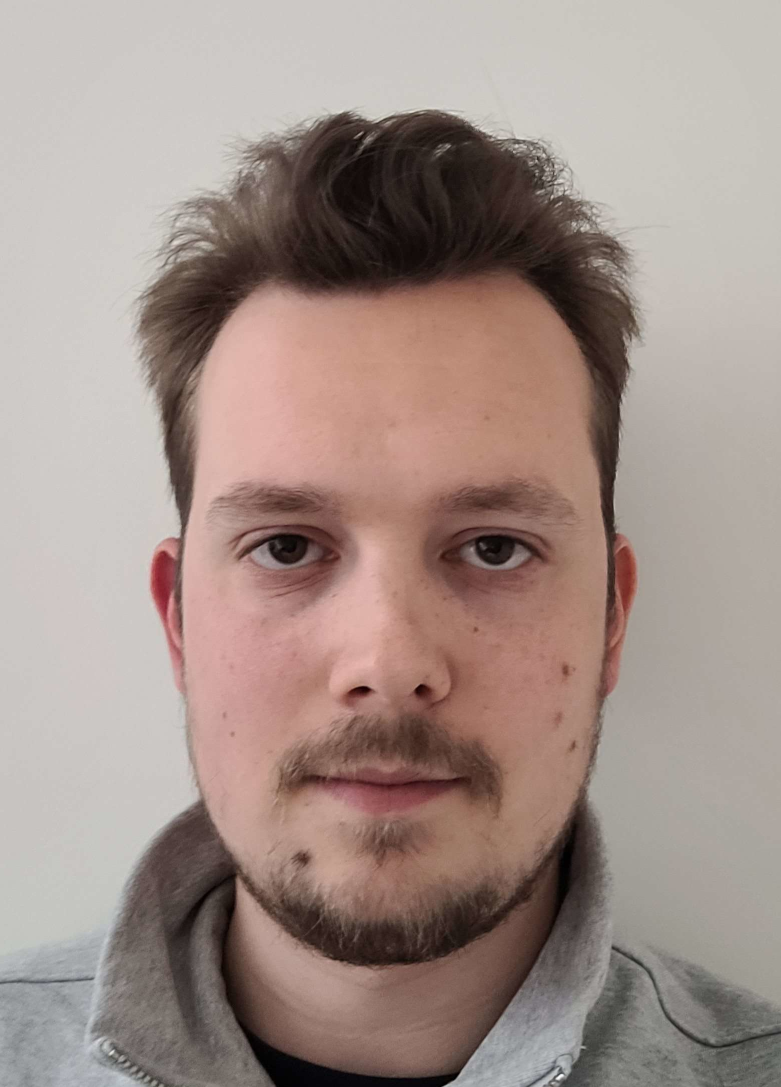}}]{Félix Riedel} is currently working towards his PhD. degree on digital in-line holography, deep learning and inverse problems. He received the signal processing engineering degree from Télécom Saint-Étienne and the M.Sc. degree in image processing from Jean Monnet university in 2024. His research interests include deep/machine learning for image reconstruction and signal processing.
\end{IEEEbiography} 

\begin{IEEEbiography}[{\includegraphics[width=1in,height=1.25in,clip,keepaspectratio]{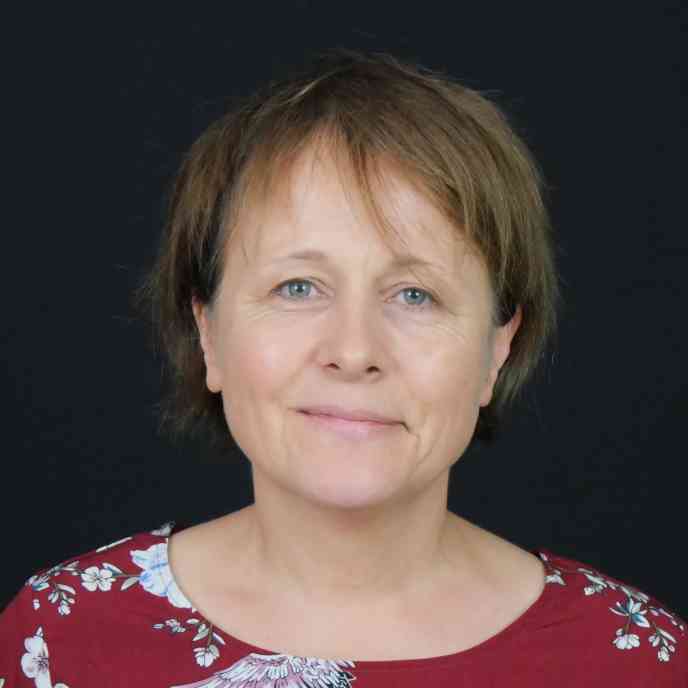}}]{Corinne Fournier} received the Ph.D. degree in image science from Université Jean Monnet (France) in 2003. She is currently Associate Professor with the Laboratoire Hubert Curien, a joint research unit of CNRS, Université Jean Monnet. She obtained her Habilitation à Diriger des Recherches (HDR) in 2017. Her research interests include computational imaging, design of optical-digital imaging systems, inverse problems for image reconstruction, digital holography, physics informed machine learning and biomedical applications. 
\end{IEEEbiography}

\begin{IEEEbiography}[{\includegraphics[width=1in,height=1.25in,clip,keepaspectratio]{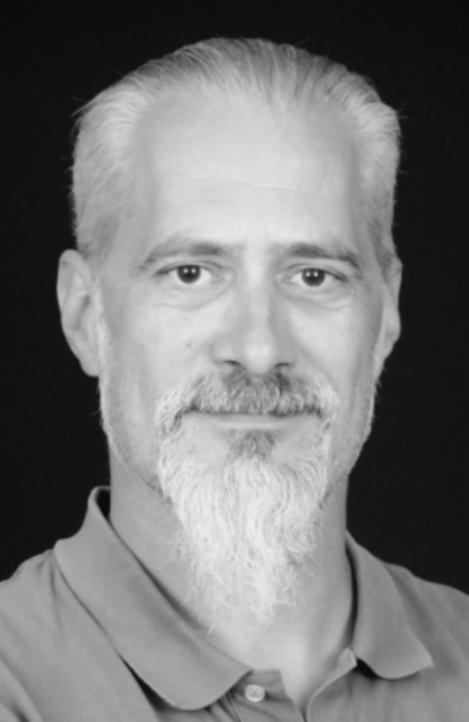}}]{Thomas Olivier}
graduated from Ecole Centrale Marseille and received his Ph.D. degree in optics, electronics, optronics and systems from University of Aix-Marseille in 2004. He worked for Institut Fresnel, Marseille from 1998 to 2004. Since 2005, he is an assistant professor at Jean Monnet University, in Laboratoire Hubert Curien, UMR CNRS 5516, Saint-Etienne, France. After he worked on various biomedical applications of microscopy, he joined the ODIR team (Optical Design and Image Reconstruction) on digital holographic microscopy projects. His research interests concern computational microscopy and more particularly holography for biomedical applications.
\end{IEEEbiography}

\begin{IEEEbiography}[{\includegraphics[width=1in,height=1.25in,clip,keepaspectratio]{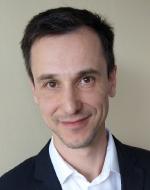}}]{Lo\"ic Denis}
(Senior Member, IEEE) received the M.Sc. degree from CPE Lyon, Villeurbanne, France, in 2003, and the Ph.D. degree from the Université de Saint-Etienne, Saint-Étienne, France, in 2006.
He is currently a Full Professor with the Université Jean Monnet Saint-Etienne, Saint-Étienne. His research interests include image denoising and reconstruction, robust signal processing, source detection, and machine learning with applications in remote sensing, diffractive microscopy, and astronomy.
Dr. Denis was a co-recipient of the IEEE ICIP Best Student Paper Award in 2010, the EUSPICO Best Student Paper Award in 2015, the IEEE Geoscience and Remote Sensing Society 2016 Transactions Prize Paper Award in 2016, and the 2021 IEEE GRSS Symposium Prize Paper Award. He served as an Associate Editor for the IEEE TRANSACTIONS ON IMAGE PROCESSING.
\end{IEEEbiography}

\newpage
\includepdf[pages=-]{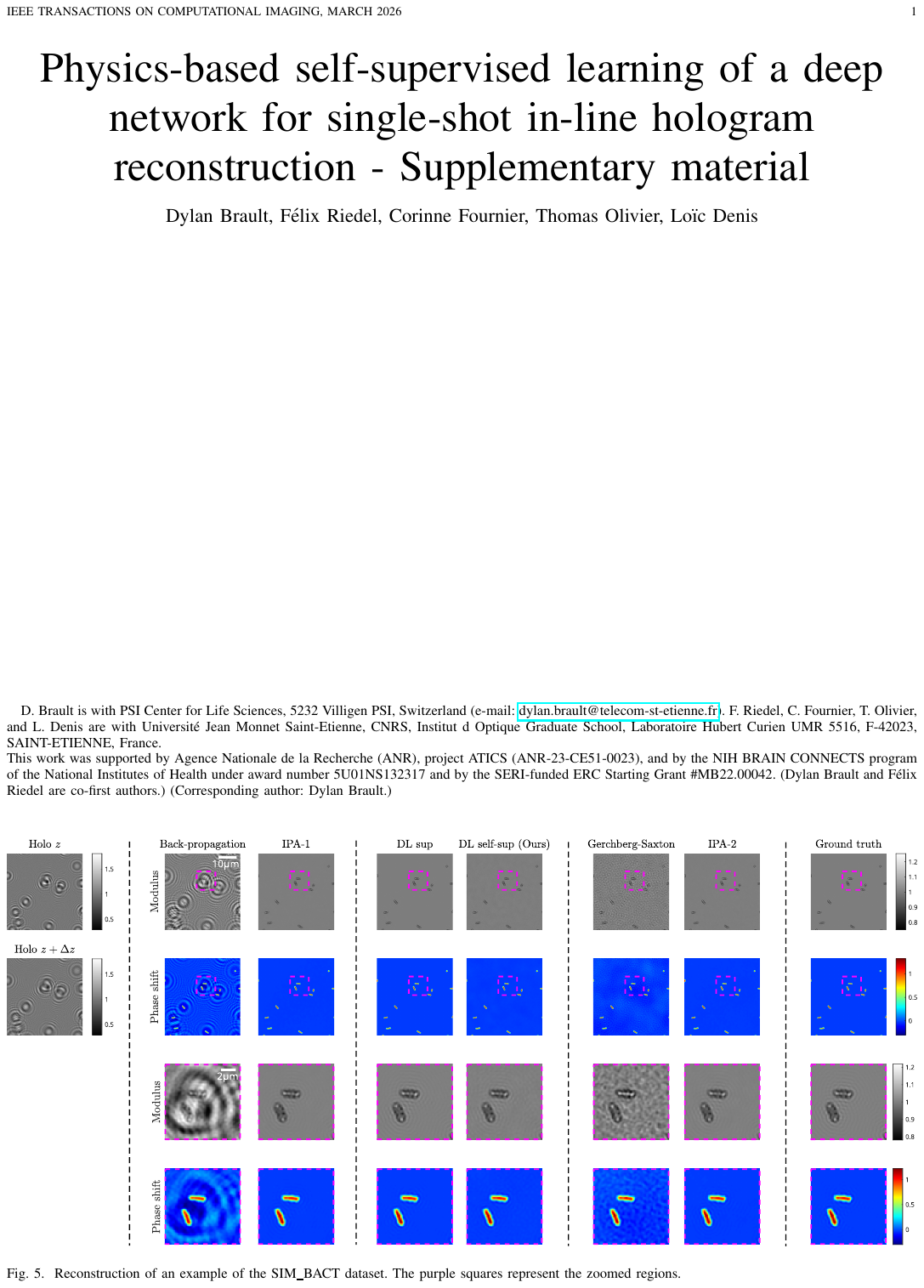}

\end{document}